\documentclass[12 pt]{article}
\usepackage{amssymb,amsthm,color,graphics}
\usepackage{amsmath}

\theoremstyle{plain}
\newtheorem{theorem}{Theorem}

\newtheorem{prop}[theorem]{Proposition}

\theoremstyle{definition}
\newtheorem{definition}{Definition}

\newcommand{\ket}[1]{\ensuremath{\left|#1\right>}}
\newcommand{\bra}[1]{\ensuremath{\left<#1\right|}}
\newcommand{\braket}[2]{\ensuremath{\left<#1|#2\right>}}

\newcommand{\ketbra}[2]{| #1 \rangle\langle #2 |}

\newcommand{\myket}[1]{{|#1\rangle}}

\newcommand{\myketbra}[2]{| #1 \rangle\langle #2 |}

\newcommand{\be}{\begin{equation}}
\newcommand{\ee}{\end{equation}}
\newcommand{\bc}{\begin{center}}
\newcommand{\ec}{\end{center}}
\newcommand{\bea}{\begin{eqnarray}}
\newcommand{\eea}{\end{eqnarray}}

\newcommand{\tr}{\text{Tr}}

\def\opone{\leavevmode\hbox{\small1\kern-3.8pt\normalsize1}}

\newcommand{\sys}[1]{{#1}}

\newcommand{\setft}[1]{\mathrm{#1}}
\newcommand{\lin}[1]{\setft{L}\left(#1\right)}

\newcommand{\pos}[1]{\setft{Pos}\left(#1\right)}

\def\I{\mathbb{I}}
\def\({\left(}
\def\){\right)}
\def\X{\mathcal{X}}
\def\Y{\mathcal{Y}}

\begin{document}

\bc {\LARGE \bf Public-key cryptography based on bounded quantum
reference frames}

\medskip
\renewcommand{\thefootnote}{\fnsymbol{footnote}}
{\footnotesize L. M. Ioannou$^{1,}$\footnote{lmi@iqc.ca} and M.
Mosca$^{1,2,}$\footnote{mmosca@iqc.ca} \\ {\it
$^1$Institute for Quantum Computing, University of Waterloo, \\
200 University Avenue, Waterloo, Ontario, N2L 3G1, Canada \\
$^2$Perimeter Institute for Theoretical Physics\\31 Caroline Street
North, Waterloo, Ontario, N2L 2Y5, Canada}} \ec

\bibliographystyle{unsrt}
\begin{quote}{\small
We demonstrate that the framework of bounded quantum reference
frames has application to building quantum-public-key cryptographic
protocols and proving their security. Thus, the framework we
introduce can be seen as a public-key analogue of the framework of
Bartlett et al. \cite{BRS04}, where a private shared reference frame
is shown to have cryptographic application. The protocol we present
in this paper is an identification scheme, which, like a digital
signature scheme, is a type of authentication scheme. We prove that
our protocol is both reusable and secure under the honest-verifier
assumption.  Thus, we also demonstrate that secure reusable
quantum-public-key authentication is possible to some extent.
}\end{quote}

\section{Introduction}

Since its inception, the focus of quantum cryptography has been on the
symmetric-key model, where Alice and Bob attempt to generate private shared correlations (as in quantum key distribution \cite{BB84,Eke91}) or
are assumed to hold them (as in quantum authentication \cite{BCGST02}).  Such correlations
can usually be defined or encoded by a string of bits---the secret
key---but Bartlett et al. \cite{BRS04} showed that they may also
take the form of a private shared reference frame.  Symmetric-key
quantum protocols are usually unconditionally secure, meaning that
the sole assumption is that (some part of) quantum theory is
correct; however, Damgaard et al. \cite{DFSS05,DFSS07} have
investigated information-theoretically secure protocols, such as password-based identification and bit commitment, in the
bounded quantum storage model, where an extra assumption is that the size or
quality of the adversary's quantum memory is limited (see also Refs
\cite{Sch07, KWW09, Sch10}).\footnote{The bounded storage model for classical protocols (e.g. Ref. \cite{CM97}), where the adversary's classical memory is assumed to be bounded, also gives information-theoretic security.}

Going beyond the symmetric-key model, but retaining unconditional
security, Gottesman and Chuang \cite{qphGC01} introduced
quantum-public-key cryptography---where the public keys are quantum
systems, each of whose state encodes the (same) classical private
key---by giving a secure one-time (digital) signature scheme for
signing classical messages.

A public-key framework eliminates the need for Alice and Bob to
establish private shared correlations, which has practical
advantages in large networks of users (where there may be many
``Alices'' or ``Bobs'').  Alice chooses a random private
key, creates copies of the corresponding public key, and distributes
the copies in an authenticated fashion to all potential ``Bobs''. In
principle, this asymmetric setup allows, e.g., any Bob to send
encrypted messages to Alice or to verify any signature for a message
that Alice digitally signed, thus significantly reducing the number
of secret/private keys involved as compared to the case where each
Alice-Bob pair shares a secret key and uses symmetric-key protocols.
Thus the public-key framework vastly simplifies key distribution,
which is often the most costly part of any cryptosystem.  Note that
the security of classical public-key protocols is necessarily based
on computational assumptions \cite{MvOV96}.

The mapping that takes a private key to the state of the
corresponding quantum public key is always assumed to be publicly
known.  Furthermore, in any reasonable quantum-public-key system,
the states of two quantum public keys corresponding to two different
private keys always have overlap less than $(1-\delta)$, for some
positive and publicly known $\delta$.  Thus, a striking aspect of
quantum-public-key cryptography that sets it apart from its
classical counterpart is that the number of copies of the (quantum)
public key in circulation must be limited. If this were not the
case, then an adversary could collect an arbitrarily large number of
copies, measure them all, and determine the private key.

The limit on the number of copies of the quantum public key implies
that not everyone can use the protocol; however, in practice, the
maximum number of users (or uses) of any particular protocol can be
estimated, and thus the parameters of the protocol can be adjusted
so that the limit allows for this maximum. Increasing this limit
would presumably result in a less efficient instance of the
protocol, and this is one kind of tradeoff between efficiency and
usability in the quantum-public-key setting. Another kind concerns
reusability. For instance, the abovementioned signature scheme is
``one-time'' because only one message may be signed under a
particular key-value, even though many different users can verify
that one signature. If a second message needs to be signed, the
signer must choose a new private key and then distribute
corresponding new public keys. One open problem is thus whether
there exist reusable signature schemes, where either the same copy
of the public key can be used to verify many different
message-signature pairs securely, or where just the same key-values
can be used to verify many different message-signature pairs
securely (but a fresh copy of the public key is needed for each
verification). The latter notion of ``reusability'' is what we adopt
here.

What makes a key \emph{public}?  In principle, Alice's
public-key-generation algorithm, which takes as input the private
key and outputs one copy of the quantum public key, may output a
system in a pure state or a mixed state, from Alice's point of view
(a mixed state is a fixed probabilistic distribution of pure
states).  In the original framework of Gottesman and Chuang, the
algorithm is assumed to produce a system in a pure state.  For some
applications, like digital signature schemes, this purity is
crucial; for, otherwise, Alice could cheat by sending different
public keys to different ``Bobs''. Purity prevents Alice's cheating
here because different ``Bobs'' can compare their copies of the
public key via a ``distributed \textsc{swap}-test'' \cite{qphGC01}
to see if they are the same (with high probability), much like can
be done in the case of classical public keys.  But the ability to do
an equality test benefits any scheme, since an adversary who tries
to substitute bad keys for legitimate ones could thus be caught.
Indeed, if the public-key-generation algorithm produces a mixed
state, since there is no equality test guaranteed to recognize when
two mixed states are equal, then no such test for equality of public
keys may be possible---this is at odds with what it means to be
``public'', i.e., publicly verifiable.\footnote{Other authors have
defined the framework to include mixed public keys, and Ref.
\cite{KKNY05} proposes an encryption scheme with mixed public keys
that is reusable and unconditionally secure \cite{HKK08}.}~~While
the scheme we present in this paper does not explicitly make use of
the ``distributed \textsc{swap}-test'' (since we assume the public
keys have been distributed securely), it can do so in principle.  We
view this as analogous to how modern public-key protocols do not
explicitly specify an equality test among unsure ``Bobs'', but how
the framework naturally allows such a test which would thwart
attempts to distribute fake public keys.


Our work appears to be of a dramatically different character when
compared to other explorations of quantum-public-key protocols
\cite{Got05,KKNY05,HKK08,Kak06,Nik08}:  we demonstrate that the
framework of bounded quantum reference frames \cite{BRST06} has
application to building such protocols and proving their security.
Thus, the framework we introduce can be seen as a public-key
analogue of the framework of Bartlett et al. \cite{BRS04}.

We stress that our work in public-key quantum cryptography strives
for unconditional security, as opposed to security based on
computational assumptions \cite{MvOV96}. In particular, our work is
unrelated to the work in Ref. \cite{OTU00}, where classical
public-key systems (whose security must be based on computational assumptions) are constructed that require a quantum computer
for the generation of the public keys.

The protocol we present in this paper is an identification scheme,
which, like a digital signature scheme, is a type of authentication
scheme. Authentication schemes are not concerned with ensuring the
\emph{privacy} of information, but rather seek to ensure its
\emph{integrity}.  For example, digital signature schemes (and
message authentication codes) ensure the integrity of origin of
messages, whereas identification schemes ensure the integrity of
origin of communication in real time \cite{MvOV96}.  Identification
protocols are said to ensure ``aliveness''---that the entity proving
its identity is active at the time the protocol is executed.

We prove that our identification protocol is both reusable and
secure under the honest-verifier assumption (defined in the next
section). Thus, we also demonstrate that secure reusable
quantum-public-key authentication is possible to some extent.

We now proceed with a description of our protocol (Section
\ref{idscheme}) and the honest-verifier security proof (Section
\ref{sec_security}).

\section{An identification scheme}\label{idscheme}

In the following, Alice and Bob are always assumed to be honest
players and Eve is always assumed to be the adversary.  Suppose
Alice generates a private key and authentically distributes copies
of the corresponding public key to any potential users of the
scheme, including Bob.

The following is a description (adapted from Section 4.7.5.1 in
Goldreich's book \cite{Gol01}) of how a secure public-key identification scheme
works.  If Alice wants to identify herself to Bob (i.e. prove that
it is she with whom he is communicating), she invokes the
identification protocol by first telling Bob that she is Alice, so
that Bob knows he should use the public key corresponding to Alice
(assuming Bob possesses public keys from many different people). The
ensuing protocol (whatever it is) has the property that the
\emph{prover} Alice can convince the \emph{verifier} Bob (except,
possibly, with negligible probability) that she is indeed Alice, but
an adversary Eve cannot fool Bob (except with negligible
probability) into thinking that she is Alice, even after having
listened in on the protocol between Alice and Bob or having
participated as a (devious) verifier in the protocol with Alice
several times. An \emph{honest-verifier identification protocol} is
only intended to be secure under the extra assumption that, whenever
Eve engages the prover Alice in the protocol, Eve follows the
verification protocol as if she were honest.  Note that no
identification protocol is secure against an attack where Eve
concurrently acts as a verifier with Alice and as a prover with Bob (but note also that, in such a case, the ``aliveness'' property is still guaranteed).
Note also that, by our definition of ``reusable,'' an identification
scheme is considered reusable if Alice can prove her identity many
times using the same key-values but the verifier needs a fresh copy
of the public key for each instance of the protocol.

A couple of remarks are in order:
\begin{itemize}
\item In practice, public-key identification schemes are implemented in
smart-card systems (e.g., inside an automated teller machine (ATM)
for access to a bank account, or beside a doorway for access to a
building), so that the smart card ``proves'' to the card reader that
it is authorized.\footnote{Note that it is not a user's personal
identification number (PIN) that functions as the prover's private
key; the PIN only serves to authenticate the user to the smart card
(not the smart card to the card reader).}~ In such situations, it
may be relatively difficult for an adversary to tamper with the
verification procedure that is encoded in the card reader, in which
case an honest-verifier identification protocol may suffice. For
example, an honest-verifier identification protocol is secure
against the class of attacks whereby an adversary collects the
maximum number of legitimate copies of the public key and uses these
in conjunction with a phony smart card to act as a dishonest prover.

\item Note that public-key identification can be trivially achieved via a
digital signature scheme (Alice signs a random message presented by
Bob), but since we do not know of an unconditionally secure and
reusable digital signature scheme, our scheme is noteworthy.
Similarly, public-key identification can be achieved with a
public-key encryption scheme (Bob sends an encrypted random
challenge to Alice, who returns it decrypted), but we do not know of
an unconditionally secure and reusable public-key encryption scheme
(that uses pure public keys; though, see Ref. \cite{Got05} for a
promising candidate).
\end{itemize}

A summary of our protocol is as follows. Alice chooses a private
phase reference and distributes a limited number of samples of her
reference frame as quantum public keys. The samples are used by Bob
to verify that the prover is actually Alice. Because Alice has a
perfect phase reference, she can carry out the identification
protocol with no error (assuming perfect quantum channels). But,
because Eve only has a bounded quantum phase reference frame (in the
form of a limited number of copies of the public key), she
inevitably incurs an error that Bob can detect with sufficiently
high probability (we discuss quantum phase reference frames
in more detail in Section \ref{sec_PhaseRFs}).  

\subsection{Protocol specification}\label{sec_ProtSpec}

Our identification protocol takes the form of a typical
``challenge-response'' interactive proof system, consisting of a kernel (or
subprotocol) that is repeated several times in order to amplify the
security, i.e., reduce the probability that an adversary can break
the protocol.  We assume all quantum channels are perfect.

\subsubsection*{Parameters}

\begin{itemize}
\item The \emph{security} parameter $s \in \mathbb{Z}^{+}$

\subitem $\diamond$ equals the number of kernel iterations.

\subitem $\diamond$ The probability that Eve can break the protocol
(in an honest-verifier setting) is exponentially small in $s$.

\item The \emph{reusability} parameter $r \in \mathbb{Z}^{+}$

\subitem $\diamond$ equals the maximum number of copies of the
quantum public key in circulation and

\subitem $\diamond$ equals the maximum number of times the protocol
may be executed by Alice, before she needs to pick a new private
key.
\end{itemize}

\subsubsection*{Keys}

\begin{itemize}
\item The \emph{private key} is
\begin{eqnarray}\label{private_key}
(x_1,x_2,\ldots,x_s),
\end{eqnarray}

where Alice chooses each $x_j$, $j=1,2,\ldots,s$, independently and
uniformly randomly from $\{1,2,\ldots,2r+1\}$.

\subitem $\diamond$ The value $x_j$ is used only in the $j$th
kernel-iteration.

\item One copy of the \emph{public key} is an $s$-partite system in the state
\begin{eqnarray}\label{public_key}
\sys{\otimes_{j=1}^s \myket{\psi_{x_j}}},
\end{eqnarray}
where (omitting normalization factors)
\begin{eqnarray}\label{public_key}
\myket{\psi_{x_j}} := \ket{0} + e^{2\pi i x_j /(2r+1)}\ket{1}.
\end{eqnarray}

\subitem $\diamond$ Alice authentically distributes (e.g. via
trusted courier) at most $r$ copies of the public key.

\subitem $\diamond$ The $j$th subsystem of the public key (which
 is in the state $\sys{\myket{\psi_{x_j}}}$) is only used in the $j$th
 kernel-iteration.

\end{itemize}

\subsubsection*{Actions}

\begin{itemize}
\item The \emph{kernel} $\mathcal{K}(x)$ of the protocol is the following three steps,
where we use the shorthand
\begin{eqnarray}
\phi_x := 2 \pi x / (r+1),
\end{eqnarray}
and where we have dropped the subscript ``$j$'' from ``$x_j$''.

\subitem (1) Bob creates $\sys{\ket{0}\ket{1} + \ket{1}\ket{0}}$,
and sends one register of this system to Alice.

\subitem (2) Alice measures the received register in the basis
$\{\myket{0} \pm e^{i \phi_x}\myket{1}\}$. If the state of the
register immediately after the measurement is $\myket{0}+e^{i
\phi_x}\myket{1}$, then Alice sends ``0'' to Bob; otherwise, Alice
sends ``1''.

\subitem (3) If Bob receives ``1'', then he applies the Pauli-$Z$
gate
\begin{eqnarray}
Z := \left[\begin{array}{cc} 1 & 0 \\ 0 & -1
\end{array}\right]
\end{eqnarray}
to the register that he kept in Step 1.  Finally, Bob
\textsc{swap}-tests\footnote{The \textsc{swap}-test of two registers
(labelled 2 and 3) in the states $\myket{\xi}_2$ and
$\myket{\chi}_3$ is a measurement (with respect to the computational
basis $\lbrace \myket{0}_1, \myket{1}_1\rbrace$) of the control
register (labelled 1) of the state
\begin{eqnarray}
(H_1 \otimes I_2 \otimes
I_3)(c-\textsc{swap}_{2,3}){(\myket{0}_1+\myket{1}_1)}\myket{\xi}_2\myket{\chi}_3/{\sqrt{2}},
\end{eqnarray}
where $H_1$ is the usual Hadamard gate (applied to register 1) and
$c-\textsc{swap}_{2,3}$ is the controlled-\textsc{swap} gate.  The
probability that the state is $\ket{0}_1$ immediately after the
measurement---which corresponds to a \emph{pass}---is
$(1+|\braket{\xi}{\chi}|^2)/2$. When the registers 2 and 3 are in
the mixed states $\rho$ and $\rho'$, this probability is
$(1+\textrm{tr} (\rho\rho'))/2$.}~ this register with his authentic
copy of $\sys{\myket{\psi_x}}$.

\item  When Alice wants to identify herself to Bob, they take the
following actions:

\subitem (\emph{i}) Alice checks that she has not yet engaged in the
protocol $r$ times before with the current value of the private key;
if she has, she aborts (and refreshes the private and public keys).

\subitem (\emph{ii}) Alice sends Bob her purported identity
(``Alice''), so that Bob may retrieve the public keys corresponding
to Alice.

\subitem (\emph{iii}) The kernel $\mathcal{K}(x)$ is repeated $s$
times, for $x=x_1, x_2,\ldots, x_s$.  Bob ``accepts'' if all the
\textsc{swap}-tests passed; otherwise, Bob ``rejects''.

\end{itemize}

\subsection{Completeness of the protocol}

It is clear that the protocol is correct for honest players, that
is, Bob always ``accepts'' when Alice is the prover. To see this,
note that, up to global phase, the state $\ket{0}\ket{1} + \ket{1}\ket{0}$
equals
\begin{eqnarray}\label{bell}
(\myket{0}+e^{i \phi_x}\myket{1})(\myket{0}+e^{i \phi_x}\myket{1}) -
(\myket{0}-e^{i \phi_x}\myket{1})(\myket{0}-e^{i \phi_x}\myket{1}).
\end{eqnarray}
\noindent In the next section, we will prove that the protocol is also secure, under the honest-verifier assumption.

\section{Honest-verifier security}\label{sec_security}

Let us clearly define what Eve is allowed to do in our
honest-verifier model.  Eve can
\begin{itemize}
\item passively monitor Alice's and Bob's interactions (which means
that Eve can read the classical bits sent by Alice, and read the bit
that indicates whether Bob ``accepts'' or ``rejects''), and

\item participate as the verifier, but only performing the
actions as if she were honest, and

\item participate as the prover in one or more instances of the protocol.
\end{itemize}
Note that Eve is assumed not to be able to actively interfere with
Alice's and Bob's communications when Alice and Bob are
participating in the protocol, as this would allow Eve to be a
dishonest verifier by stealing the qubits Bob sends to Alice and
replacing them with her own qubits.

Immediately, we see that Eve's passive monitoring only gives her
independent and random bits (plus the bit corresponding to ``accept''), and thus
gives her no useful information (in the sense that she may as well
generate random bits herself). We can therefore ignore the effects
of her passive monitoring.

With regard to Eve acting as verifier, note that, in the kernel
iteration $\mathcal{K}(x)$, Eve can at best extract one extra copy
of $\myket{\psi_{x}}$ from Alice when Eve follows the verifier
protocol honestly in Step 1, by not bothering to do the
\textsc{swap}-test in Step 3.  Eve is technically not
allowed not to do the \textsc{swap}-test in our honest-verifier
setting, but we show that---even if she is allowed---then the
protocol is secure (as long as Eve is honest in Step 1). This allows Eve to obtain a maximum of $r$ extra
copies (recall Alice only participates in the protocol $r$ times
before refreshing her keys) of the public key in addition to any
copies she obtains legitimately. Let $t$ be the total number of
copies of the public key that Eve has in her possession. Note that
$t \leq 2r-1$, since we always assume that at least one copy is left for Bob,
so that Eve can carry out the protocol with him.

Therefore, to prove security in our setting, it suffices to consider
attacks where Eve is armed with her $t$ copies of the public key and
she participates as a prover in order to try to cause Bob to
``accept''. We use the following definition of ``security''.

\begin{definition}[Security]\label{def_Security}
An honest-verifier identification protocol (for honest prover Alice
and honest verifier Bob) is \emph{secure with error $\epsilon$} if
the probability that Bob ``accepts'' when any adversary Eve
participates in the protocol as a prover is less than $\epsilon$
(assuming that, whenever Eve engages Alice in the protocol, Eve
follows the verification protocol honestly).
\end{definition}

We will assume that Eve has always extracted the $r$ illegitimate
copies of the public key from Alice, and we define $t'$ to be the
number of copies that Eve obtained legitimately:
\begin{eqnarray}
t = r + t'.
\end{eqnarray}
Note that Eve can make at most $(r-t')$ attempts at fooling Bob,
i.e., causing Bob to ``accept''.  Most of the argument, beginning in
Section \ref{sec_indiv_attacks}, is devoted to showing that
\begin{eqnarray}\label{beginning}
&&\textrm{Pr}[\textrm{Eve fools Bob on first attempt, using $t$
copies}]
\\\label{beg2} &\leq& (1-1/8(t+1))^s.
\end{eqnarray}
In general, Eve learns something from one attempt to the next;
however, because Eve can simulate her interaction with Bob at the
cost of using one copy of $\myket{\psi_x}$ per simulated iteration of $\mathcal{K}(x)$, we
have, for $\ell=2,3,\ldots,(r-t')$,
\begin{eqnarray}\nonumber
&&\textrm{Pr}[\textrm{Eve fools Bob on $\ell$th attempt, using $t$
copies}]\\\nonumber &\leq&\textrm{Pr}[\textrm{Eve fools Bob on first
attempt, using $(t+\ell-1)$ copies}].
\end{eqnarray}
Given this, we use the union bound:
\begin{eqnarray}\nonumber
&&\textrm{Pr}[\textrm{Eve fools Bob at least once, using $t$
copies}] \\
\nonumber &\leq& \sum_{\ell=1}^{r-t'} \textrm{Pr}[\textrm{Eve
fools Bob on $\ell$th attempt, using $t$ copies}]\\
\nonumber &\leq& \sum_{\ell=1}^{r-t'} \textrm{Pr}[\textrm{Eve fools
Bob on first attempt,
using $(t+\ell-1)$ copies}]\\
\nonumber &\leq& \sum_{\ell=1}^{r-t'}
(1-1/8(t+\ell))^s\\
\nonumber &\leq& (r-t')(1-1/16r)^s.
\end{eqnarray}
It follows that the probability that Eve can fool Bob at least once,
that is, break the protocol, is
\begin{eqnarray}\label{eqnPbreak}
P_{\textrm{\footnotesize break}} \leq r(1-1/16r)^s,
\end{eqnarray}
which, for fixed $r$, is exponentially small in $s$.  Note that this
bound is likely not tight, since it ultimately assumes that all of Eve's
attempts are equally as powerful.  In particular, this bound assumes
that Eve's copies do not degrade with each use.  A more detailed
analysis using results about degradation of quantum reference frames
 \cite{BRST06} may be possible.

From Eq. (\ref{eqnPbreak}) follows our main theorem:
\begin{theorem}[Honest-verifier-security of the protocol] For any $\epsilon >0$ and any $r \in \mathbb{Z}^+$,
the identification protocol specified in Section \ref{sec_ProtSpec}
is secure with error $\epsilon$ according to Definition
\ref{def_Security} if
\begin{eqnarray}\label{eqn_scal}
s \in \Omega(r \log(r/\epsilon)).
\end{eqnarray}
\end{theorem}
\noindent The theorem shows how the efficiency of the protocol
scales with its reusability.

The remainder of the paper establishes the bound in Lines
(\ref{beginning}) and (\ref{beg2}).

\subsection{Sufficiency of individual attacks}\label{sec_indiv_attacks}

At each iteration, Eve should take some action, which we
may assume takes the form of a measurement, in order to get an
answer to send back to Bob.  In general, Eve can mount a coherent attack,
whereby her actions during iteration $j$ may involve systems that she used or will use
in previous or future iterations as well as systems consisting of copies of the state $\myket{\psi_{x_k}}$ for any $k$---not just for $k = j$.  Intuition suggests that, since each $x_j$ is \emph{independently}
selected from the set $\{1, 2, \ldots,
2r+1\}$, Eve's
measurement at iteration $j$ may be assumed to be independent of her measurement at any other iteration and in particular does not need to act on any components of copies of
the public key other than those corresponding to the copies of $\myket{\psi_{x_j}}$.  In other words, it seems plausible that Eve's optimal strategy can without loss of generality consist of the ``product'' of identical optimal strategies for each iteration
individually.  Indeed, this intuition can be shown to be correct by combining a technique from Ref. \cite{JUW09}, for expressing the maximum probability of acceptance in a two-message quantum interactive proof system as a semidefinite program, with a result in Ref. \cite{MS07}, which implies that the semidefinite program satisfies the product rule that we need; see the Appendix for a proof.  The remainder of
Section \ref{sec_security} establishes that the probability of
passing the \textsc{swap}-test for any particular iteration is at
most $1-1/8(t+1)$, and thus, by the result proved in the Appendix, the probability of passing all $s$
\textsc{swap}-tests is at most $(1-1/8(t+1))^s$.

\subsection{Equivalence of discrete and continuous private phases}

Now, we show that, from Bob's and Eve's points of view, Alice's
choosing the private phase angle $\phi_x$ from the discrete set $\{2
\pi x/(2r+1): x=1,2,\ldots,2r+1\}$ is equivalent to her choosing the
phase angle from the continuous interval $[0,2\pi)$.  We have argued
that the only information that Eve or Bob (or anyone but Alice) has about
$\phi_x$ may be assumed to come from a number of copies of
$\sys{\myket{\psi_x}}$ that can be no greater than $2r$ (there are $r$
legitimate copies of the public key, and one can extract $r$ more
copies from Alice); let this number be $c$, where $1 \leq c \leq
2r$. We may describe the state of these $c$ systems by the density
operator
\begin{eqnarray}
\frac{1}{2r+1}\sum_{x = 1}^{2r+1} \frac{1}{2^c}((\ket{0} + e^{2 \pi
i x/(2r+1)}\ket{1})(\bra{0} + e^{-2 \pi i
x/(2r+1)}\bra{1}))^{\otimes c}.
\end{eqnarray}
Had $\phi_x$ been chosen uniformly from $\{2 \pi x/(2r+1): x\in
[0,2r+1)\} = [0,2\pi)$, they would describe the state by
\begin{eqnarray}
\frac{1}{2\pi}\int_0^{2 \pi} \frac{1}{2^c}((\ket{0} + e^{i
\phi}\ket{1})(\bra{0} + e^{-i \phi}\bra{1}))^{\otimes c} d\phi.
\end{eqnarray}
It is straightforward to show that the above two density operators
are both equal to
\begin{eqnarray}\label{symm}
\frac{1}{2^c}\sum_{w=0}^c {c \choose w} \myketbra{S^c_w}{S^c_w},
\end{eqnarray}
where $\myket{S^c_w}$ is the normalized symmetric sum of all $c
\choose w$ states in $\{\myket{0}, \myket{1}\}^{\otimes c}$ whose
binary labels have Hamming weight $w$.\footnote{This requires the
following two facts: (1) for any integer $a$,
\begin{eqnarray}
\frac{1}{2\pi}\int_{0}^{2\pi}  e^{i a \theta} d\theta= \left\{
\begin{array}{ll}
                0 & \mbox{ if $a\neq 0$ }, \\
                1 & \mbox{ otherwise     };
                \end{array}
                \right.
\end{eqnarray}
and (2) for any integer $p\geq 2$ and integer $a$:
\begin{eqnarray}
\frac{1}{p}\sum_{k=1}^p e^{2 \pi i a k/p} =\left\{
\begin{array}{ll}
                0 & \mbox{ if $a$ is not a multiple of $p$}, \\
                1 & \mbox{ otherwise     },
                \end{array}
                \right.
\end{eqnarray}
where the second fact is applied at $p=2r+1$.} Thus, without loss,
we may drop the subscript ``$x$'' on ``$\phi_x$'', write ``$\phi$''
for Alice's private phase angle, and assume she did (somehow) choose
$\phi$ uniformly randomly from $[0,2\pi)$.\footnote{One way to
interpret this result is that even if Alice encodes infinitely many
bits into $\phi$, it is no better than if she encoded $\lceil
\log_2(2r+1) \rceil$ bits. Note that if Eve performs an optimal
phase estimation \cite{vDdAEMM07} in order to learn $\phi$ and then
cheat Bob, she can only learn at most $\lfloor \log_2(2r-1)\rfloor$
bits of $\phi$ (here, we assume Eve has $2r-1$ copies of the public
key, having left Bob one copy), whereas Alice actually encoded
$\lceil \log_2(2r+1) \rceil$ bits into $\phi$.}

\subsection{Sufficiency of maximizing successful guessing probability}

The security of our protocol follows from a result of Bartlett et
al. \cite{BRST06}, which concerns a slightly different problem for
Eve than the problem of her trying to cheat (fool) Bob.  This
different problem is for Eve, using her $t$ copies of $\ket{0}+e^{i \phi}\ket{1}$,
to guess whether she has been given the
system $\sys{\ket{0} + e^{i \phi}\myket{1}}$ or $\sys{\myket{0} - e^{i
\phi}\myket{1}}$, where each case occurs with equal probability and
$\phi$ is unknown and uniformly randomly chosen from $[0, 2\pi)$.
The purpose of this section is to show that any good cheating
strategy gives a good guessing strategy; we will show that an upper
bound on the average successful guessing probability gives an upper
bound on the cheating probability, so that, in order to prove
security, it suffices to show that the maximum successful guessing
probability is sufficiently small (which we will do in the next
section).

Any cheating strategy of Eve can be modeled as follows.  Let
\begin{eqnarray}
\myket{\pm} := \myket{0} \pm e^{i \phi}\myket{1}.
\end{eqnarray}
Recall that Bob creates a system in the state $\myket{0}\myket{1}
+\myket{1}\myket{0}$, which equals ${\myket{+}\myket{+} - \myket{-}\myket{-}}$
up to global phase. Eve's system before Bob sends one of his
registers can be represented by $\sys{\myket{\Xi}}$, which consists of
the $t$ copies of $\sys{\myket{+}}$ as well as any ancillary registers
(which we can assume are in a pure state).  Eve's
(optimal) measurement can thus be modeled by a unitary operation
$U_E$ acting on her system (labeled $E$, which now includes the
qubit Bob sends), which transforms the state of the total system as
follows:
\begin{eqnarray}\label{U}
&&\frac{1}{\sqrt{2}}{(\myket{+}_B\myket{+}_E -
\myket{-}_B\myket{-}_E)}\myket{\Xi}_E \mapsto^{U_E}\\
&&\hspace{20mm}\frac{1}{\sqrt{2}}\left(\myket{+}_B(\alpha\myket{0}_E\myket{\psi^+_0}_E
+
\beta\myket{1}_E\myket{\psi^+_1}_E) -\right.\\
&&\hspace{30mm}\left.\myket{-}_B(\gamma\myket{0}_E\myket{\psi^-_0}_E +
\delta\myket{1}_E\myket{\psi^-_1}_E)\right),
\end{eqnarray}
so that the leftmost register of Eve's system encodes her
measurement outcome.  Bob's application of the $Z$ gate conditioned
on the value of the measurement outcome can be modeled by a
controlled-$Z$ gate (where Bob's kept qubit, labeled $B$, is the
target-qubit and the leftmost qubit of Eve's system is the
control-qubit), which will take the state of the total system to
\begin{eqnarray}
\frac{1}{\sqrt{2}}\left(\myket{+}_B(\alpha \myket{0}_E
\myket{\psi^+_0}_E -
\delta\myket{1}_E\myket{\psi^-_1}_E) \right.+\\
\hspace{20mm}\left.\myket{-}_B(\beta\myket{1}_E\myket{\psi^+_1}_E -
\gamma \myket{0}_E\myket{\psi^-_0}_E)\right).
\end{eqnarray}
Let $\tau$ represent the density operator for this state after Eve's
system has been traced out.  The probability that Bob's
$\textsc{swap}$-test passes is easily calculated to be
\begin{eqnarray}
P_{\footnotesize{\textrm{pass}}} &=& \frac{1 + \bra{+}\tau\myket{+}}{2}\\
&=& \frac{1 + (|\alpha|^2 + |\delta|^2)/2}{2}.
\end{eqnarray}

Now, suppose Eve is faced with the different problem of guessing
whether Bob gave her $\myket{+}$ or $\myket{-}$, where each case occurs
with probability $1/2$ (and where $\phi$ is unknown and uniformly
random in $[0, 2\pi)$). Since, as can be seen from the mapping in
Line (\ref{U}), $U_E$ maps
\begin{eqnarray}
\myket{+}_E\myket{\Xi}_E &\mapsto^{U_E}& \alpha \myket{0}_E
\myket{\psi^+_0}_E +
\beta\myket{1}_E\myket{\psi^+_1}_E\\
\myket{-}_E\myket{\Xi}_E &\mapsto^{U_E}&
\gamma\myket{0}_E\myket{\psi^-_0}_E +
\delta\myket{1}_E\myket{\psi^-_1}_E,
\end{eqnarray}
Eve can use the same procedure she used for her attack in order to
guess which state Bob prepared: upon measuring her leftmost
register, she guesses ``$\myket{+}$'' if she gets outcome ``0'', and
otherwise she guesses ``$\myket{-}$''.  The probability that she
guesses successfully on average using this strategy is clearly
\begin{eqnarray}
P_{\footnotesize{\textrm{succ}}} &=& \frac{1}{2}\times
\textrm{Pr}(\textrm{outcome
=``0''}|\textrm{Bob prepared $\myket{+}$}) + \\
&&\hspace{10mm}\frac{1}{2}\times \textrm{Pr}(\textrm{outcome
=``1''}|\textrm{Bob
prepared $\myket{-}$})\\
&=& (|\alpha|^2 + |\delta|^2)/2.
\end{eqnarray}
Thus, any upper bound on $P_{\footnotesize{\textrm{succ}}}$ gives an
upper bound on $P_{\footnotesize{\textrm{pass}}}$.

\subsection{Bounding the successful guessing
probability}\label{sec_BoundingSuccGuessProb}

Bartlett et al. \cite{BRST06} give an expression for the average
successful guessing probability in terms of the state of a bounded
phase reference frame (which, for us, takes the form of copies of
the public key). In Section \ref{sec_PhaseRFs}, we give some
background on (bounded) phase references so that we can better
understand the result in Ref. \cite{BRST06} in order to apply it.
In Section \ref{sec_Bound}, we derive the bound on Eve's successful
guessing probability, establishing what was claimed in Lines
(\ref{beginning}) and (\ref{beg2}).

\subsubsection{Phase reference frames}\label{sec_PhaseRFs}

Consider the two qubit states ``$\myket{0} + \myket{1}$'' and ``$\myket{0}
- \myket{1}$'' (the use of quotation marks will become clear below).
Given a qubit promised to be prepared in one of those two states,
how could you decide which state the given qubit is in? In general,
without any other system to help you, you cannot, because the
question is not well defined: the states $\myket{0}$ and $\myket{1}$ are
only defined up to global phase, so, e.g., replacing $\myket{1}$ by
$-\myket{1}$ changes your answer.  A \emph{phase reference (frame)} is
a quantum-mechanical system that, when taken together with the given
qubit, fixes the relative phase between $\myket{0}$ and $\myket{1}$ in
the state ``$\myket{0} + \myket{1}$''.  This intuitive definition
suffices for our purposes.

Note that usually in quantum information processing, it is assumed
that one has a phase reference, which ascribes definite meaning to
the state ``$\myket{0} + \myket{1}$''.  Similarly, in multiparty quantum
communication/cryptography protocols, it is usually assumed that all
players involved have access to a common phase reference, so that
``$\myket{0} + \myket{1}$'' means the same thing to each party.  This is
in fact a reasonable assumption, since it has been shown that, with
modest overhead, any set of players can simulate having a common (phase) reference frame, by using the symmetric and
anti-symmetric subspaces of entangled states \cite{BRS03}.

The most popular phase reference frame occurs in optics, and is
known as a ``coherent state'', which is usually defined as the state
\begin{eqnarray}\label{eqn_coherentstate}
\myket{C_\theta} := e^{-
\frac{\alpha^2}{2}}\sum_{w=0}^{\infty}(\alpha^w /\sqrt{w!})e^{i w
\theta}\myket{w},
\end{eqnarray}
of a single (optical) mode, where $\alpha$ is a real number; this
state \emph{encodes} the \emph{(relative) phase} $e^{i \theta}$. The
number $w$ in $\myket{w}$ is the \emph{photon number}. Employing a
system in this state to first prepare the given qubit and then to
measure it, in order to solve the decision problem posed above,
fixes the relative phase of $\myket{0}$ and $\myket{1}$ in the
superposition ``$\myket{0} + \myket{1}$'' to be $e^{i\theta}$, i.e.,
this superposition is now more correctly written $\myket{0} + e^{i
\theta}\myket{1}$.\footnote{The precise way in which this preparation
and measurement works in practice is beyond the scope of this paper,
but we give an intuitive explanation. First note that it is
sufficient to implement the Hadamard gate
\begin{eqnarray}
\myket{0} &\mapsto& \myket{0} + e^{i \phi}\myket{1}\\
e^{i \theta}\myket{1} &\mapsto& \myket{0} - e^{i \phi}\myket{1}.
\end{eqnarray}
Since the coherent state is unchanged (up to global phase) under the
operation $\myket{w} \mapsto \myket{w-1}$ (i.e. the annihilation of a
photon, which we note is not a unitary operation), it is possible to
approximate the Hadamard gate by approximately mapping
$\myket{C_\theta}\myket{0} \mapsto \myket{C_\theta}(\myket{0} + e^{i
\theta}\myket{1})$ by taking a photon from the coherent state to use
in the right-hand qubit; similarly, the operation
$\myket{C_\theta}\myket{1} \mapsto \myket{C_\theta}(e^{-i \theta}\myket{0} -
\myket{1})$ may be approximated. The quality of the approximation
depends on the total energy of the coherent state. See Ref.
\cite{qphIM08} for a complete analysis of a similar task.}~
Normally, one would redefine $\myket{1}$ as $\myket{1}:= e^{i
\theta}\myket{1}$, and write the superposition as $\myket{0} + \myket{1}$.

Note that there is still a slight problem in that $\theta$ is not
really well defined, since it could be replaced by any $\theta' \neq
\theta$ and the physics of the problem would not change.  We say
that, in the definition of the coherent state, the phase $e^{i
\theta}$ is defined relative to a \emph{hidden absolute phase
reference}, which in practice means that the actual value of
$\theta$ need not be known, but what is important is that the
relative phase $e^{ i \theta}$ stays consistent throughout all the
quantum operations.

A single mode is mathematically modeled by $\mathbb{C}^{N}$, and a
basis for this space is $\{\myket{n}:n=0,1,\ldots,N-1\}$
; in general, as for the coherent state, $N$ can equal $\infty$.
Similarly, a $k$-mode (multimode) phase reference is modeled by
$(\mathbb{C}^N)^{\otimes k}$.  It is convenient to adopt this
optics-based nomenclature (modes, photons) when discussing reference
frames, though the following results are completely general and do
not rely on optical implementations.

In practice, any phase reference frame is \emph{bounded}, meaning
that the total energy, or average total photon number, of the state
is upper-bounded. For a multimode phase reference state, this bound
may take the form of an upper bound on the number $k$ of modes
and perhaps an upper bound on the photon number of each mode.  Generally,
the higher the total energy of the phase reference frame, the better
it performs in practice, i.e., the better it maintains consistent
relative phase throughout a quantum computation.

An example of a $k$-mode bounded phase reference frame encoding the
phase $e^{i \theta}$ is a system of $k$ qubits in the state
\begin{eqnarray}\label{rfstate}
(\myket{0} + e^{i \theta}\myket{1})^{\otimes k} = \sum_{w=0}^k \sqrt{{k
\choose w}}e^{i w \theta} \myket{S^k_w},
\end{eqnarray}
where $\myket{S_w^{k}}$ is the $k$-mode state defined just after Eq.
(\ref{symm}) (see Ref. \cite{Enk05} for a detailed discussion of
such ``refbits''). For this multimode phase reference frame, the
maximum photon number is 1 for each mode.

Define the unitary re-phasing map $U(\theta)$ on
$(\mathbb{C}^{k+1})^{\otimes k}$ as the mapping
\begin{eqnarray}
\myket{w_1}\myket{w_2}\cdots \myket{w_{k}} \mapsto e^{i (w_1 + w_2 +
\cdots w_{k}) \theta}\myket{w_1}\myket{w_2}\cdots \myket{w_{k}}
\end{eqnarray}
for any $\theta \in [0, 2\pi]$ and all $w_l = 0,1,\dots,k$, for
$l=1,2,\ldots,k$. A unitary operation $V$ on
$(\mathbb{C}^{k})^{\otimes k}$ is said to be \emph{phase invariant}
if $U(\theta)V U(\theta)^\dagger = V$ for all $\theta \in [0,
2\pi]$.  If $V$ is phase invariant, one does not need any phase
reference to perform $V$; e.g., Eve could use her own phase
reference (say, a coherent state encoding the phase $e^{i
\theta_E}$) to carry out $V$ on some register, and the result would
be the same (up to global phase) as if Alice performed $V$ on the
same register using her own phase reference (a coherent state
encoding the phase $e^{i \theta_A}$, $\theta_A \neq \theta_E$); for simplicity, we have assumed that Eve's and Alice's phase references are perfect (see
Section II.B of Bartlett et al. \cite{BRS07} for more details).  We
will use the fact that $U(\theta)$ is phase invariant if and only if
it is block-diagonal with respect to subspaces of constant total
photon number (see e.g. Ref. \cite{qphIM08} for a proof).

\begin{definition}[Equivalence of phase reference frames]  Suppose
$\rho_1$ and $\rho_2$ are two states of a multimode phase reference
frame.  Then a phase reference in state $\rho_1$ and a phase
reference in state $\rho_2$ are \emph{equivalent} if there exists a
phase-invariant unitary operation $V$ such that
\begin{eqnarray}
\rho_1 = V \rho_2 V^\dagger.
\end{eqnarray}
\end{definition}

Note that there is a phase-invariant unitary transformation on
$(\mathbb{C}^{k+1})^{\otimes k}$ that maps
\begin{eqnarray}
\myket{S_w^{k}} \mapsto \myket{0}^{\otimes
(k-1)}\myket{w},\hspace{2mm}\textrm{for all $w=0,1,\ldots,k$,}
\end{eqnarray}
because this mapping may be completed on
$(\mathbb{C}^{k+1})^{\otimes k}$ to a unitary operator that is block-diagonal
with respect to subspaces of constant total photon number.
Therefore, a multimode phase reference in the state in Line
(\ref{rfstate}) is equivalent to a single-mode phase reference in
the state
\begin{eqnarray}
\sum_{w=0}^k \sqrt{{k \choose w}}e^{i w \theta} \myket{w}
\end{eqnarray}
(where we have omitted the ancilla in the state $\myket{0}^{\otimes
(k-1)}$), which we note looks like the coherent state but for the
moduli of the coefficients.

Finally, we define a special class of phase references.

\begin{definition}[Covariant family of phase reference states]  Let $U(\theta)$ be
the unitary rephasing map on $\mathbb{C}^{N}$ such that $U(\theta)\myket{w}=e^{i
w \theta}\myket{w}$ for all $\theta \in [0, 2\pi]$.  Suppose that
$\{\rho(\phi)\}_{\phi \in [0, 2\pi]}$ is a family of (single-mode)
phase reference states on $\mathbb{C}^{N}$, where $\rho(\phi)$
encodes the relative phase $e^{i \phi}$. Then $\{\rho(\phi)\}_{\phi
\in [0, 2\pi]}$ is \emph{covariant} if
\begin{eqnarray}
\rho(\phi) = U(\phi) \rho(0) U(\phi)^\dagger
\end{eqnarray}
for all $\phi \in [0, 2\pi]$.  
\end{definition}

\subsubsection{The bound}\label{sec_Bound}

First we note that the effect of Alice's selection of $\phi$ serves
to completely randomize the relative phase between $\myket{0}$ and
$\myket{1}$ in any superposition of the two states, from the point
of view of anyone other than Alice.  Thus, even though we make the
usual assumption that all players (Alice, Bob, Eve) share a common
phase reference, the protocol effectively forces Alice to have a
private phase reference, leaving the other players with maximal
ignorance (but for the information contained in the copies of the
public key) of what the ``correct'' relative phase is in each
iteration of the protocol. Therefore, Eve's $t$ copies of $\ket{0} +
e^{i \phi}\ket{1}$ may be seen as a bounded multimode phase
reference encoding the phase $e^{i \phi}$ relative to a \emph{known}
common phase reference---but for \emph{unknown} and uniformly random
$\phi \in [0, 2\pi)$.\footnote{Note that our assumption that Alice,
Bob, and Eve all share a perfect common phase reference implies that
Alice can make $r$ samples of her phase reference with no
degradation of the original phase reference.  Thus, while we are
using the theoretical framework of (bounded) quantum reference
frames from Refs \cite{BRS04, BRST06} in our analysis, our initial
assumptions are different than in those works (where the standard
assumption that everyone shares a perfect common phase reference is
usually not used).}

Bartlett et al. \cite{BRST06} prove the following theorem (rephrased
for our purposes).

\begin{theorem}[Optimal probability of successful guessing \cite{BRST06}]
Suppose $\{\rho(\phi)\}_{\phi \in [0, 2\pi]}$ is a covariant family
of single-mode phase reference states, where $\rho(\phi)$ encodes
the phase $e^{i \phi}$ (relative to a known common phase reference).
Given a single mode in the state $\rho(\phi)$, for $\phi$ unknown
and uniformly random in $[0, 2\pi)$, and a qubit in one of the two
states $\myket{0} \pm e^{i \phi}\myket{1}$, where each state occurs with
probability 1/2, the optimal probability of successfully guessing
which state the given qubit is in is
\begin{eqnarray}
P_{\footnotesize{\textrm{succ}}} &=& \frac{1}{2} +
\frac{1}{2}\sum_{m=0}^\infty \Re(\bra{m+1}\rho(0) \myket{m}).
\end{eqnarray}
\end{theorem}

We showed in the previous section that Eve's multimode phase
reference is equivalent to the single-mode phase reference in the
state
\begin{eqnarray}
\rho(\phi) := \frac{1}{2^t}\sum_{w=0}^t \sum_{w' =0}^{t} \sqrt{{t
\choose w}{t \choose w'}}e^{i (w-w') \phi}\ketbra{w}{w'}.
\end{eqnarray}
Substituting this value of $\rho(\phi)$ into the above theorem gives
\begin{eqnarray}
P_{\footnotesize{\textrm{succ}}} &=& \frac{1}{2} +
\frac{1}{2}\frac{1}{2^{t}}\sum_{m=0}^{t-1} \sqrt{{t \choose m}{t
\choose {m+1}}},
\end{eqnarray}
which we can show to be in $1 -\Omega(1/{t})$ (up to logarithmic
factors) using some simple approximations.  Cheung \cite{Che09} has
improved our asymptotic bound on this quantity by showing that
\begin{eqnarray}
\frac{1}{2^{t}}\sum_{m=0}^{t-1} \sqrt{{t \choose m}{t \choose
{m+1}}} \leq 1 - \frac{1}{2(t+1)} - \frac{1}{2^{t + 1}},
\end{eqnarray}
which implies
\begin{eqnarray}
P_{\footnotesize{\textrm{succ}}} \leq 1 - {1}/{4(t+1)}.
\end{eqnarray}

It follows that
\begin{eqnarray}
P_{\footnotesize{\textrm{pass}}} \leq 1 - {1}/{8(t+1)},
\end{eqnarray}
which we recall is an upper bound on the probability that Bob's
$\textsc{swap}$-test passes in any particular kernel-iteration, when
Eve is acting as a dishonest prover and using $t$ copies of the
public key. Thus, as we argued in Section
\ref{sec_indiv_attacks} (and the Appendix), the total probability that Eve causes all $s$ of
Bob's $\textsc{swap}$-tests to pass is
\begin{eqnarray}\nonumber
&&\textrm{Pr}[\textrm{Eve fools Bob on first attempt, using $t$
copies}]\\\nonumber &\leq& (1-1/8(t+1))^s,
\end{eqnarray}
as claimed in Lines (\ref{beginning}) and (\ref{beg2}).  This
completes the proof of security of the protocol.

\section*{Acknowledgements}

We acknowledge useful discussions with Giulio Chiribella, Daniel
Gottesman, Rob Spekkens, and, in particular, John Watrous, who gave
us the argument in the Appendix that proves that the protocol is
secure against coherent attacks.  L. M. Ioannou was supported by
EPSRC, SCALA, QuantumWorks, MITACS, and IQC.  M. Mosca was supported
by NSERC, DTO-ARO, CFI, CIFAR, Ontario-MRI, CRC, OCE, QuantumWorks
and MITACS.

\bibliographystyle{unsrt}
%
%

\begin{thebibliography}{10}

\bibitem{BRS04}
Stephen~D. Bartlett, Terry Rudolph, and Robert~W. Spekkens.
\newblock Decoherence-full subsystems and the cryptographic power of a private
  shared reference frame.
\newblock {\em Phys. Rev. A}, 70:032307, 2004.

\bibitem{BB84}
C.~H. Bennett and G.~Brassard.
\newblock Quantum cryptography: {P}ublic key distribution and coin tossing.
\newblock In {\em Proceedings of IEEE International Conference on Computers,
  Systems and Signal Processing}, pages 175--179, New York, December 1984. IEEE
  Press.

\bibitem{Eke91}
Artur~K. Ekert.
\newblock Quantum cryptography based on {B}ell's theorem.
\newblock {\em Phys. Rev. Lett.}, 67(6):661--663, 1991.

\bibitem{BCGST02}
Howard Barnum, Claude Cr{\'{e}}peau, Daniel Gottesman, Adam Smith,
and Alain
  Tapp.
\newblock Authentication of quantum messages.
\newblock In IEEE Press, editor, {\em Proc. 43rd Annual IEEE Symposium on the
  Foundations of Computer Science (FOCS '02)}, pages 449--458, 2002.

\bibitem{DFSS05}
Ivan Damgaard, Serge Fehr, Louis Salvail, and Christian Schaffner.
\newblock Cryptography in the bounded quantum-storage model.
\newblock In IEEE Press, editor, {\em Proceedings of the 46th IEEE Symposium on
  Foundations of Computer Science - FOCS 2005}, pages 449--458, 2005.

\bibitem{DFSS07}
Ivan Damgaard, Serge Fehr, Louis Salvail, and Christian Schaffner.
\newblock Secure identification and {QKD} in the bounded-quantum-storage model.
\newblock {\em {CRYPTO} 2007, Lecture Notes in Computer Science},
  4622:342--359, 2007.

\bibitem{Sch07}
Christian Schaffner.
\newblock {\em Cryptography in the bounded-quantum-storage model}.
\newblock PhD thesis, University of Aarhus, 2007.
\newblock Ph.D. thesis.

\bibitem{KWW09}
Robert Koenig, Stephanie Wehner, and Juerg Wullschleger.
\newblock Unconditional security from noisy quantum storage, 2009.
\newblock arXiv:0906.1030.

\bibitem{Sch10}
Christian Schaffner.
\newblock Simple protocols for oblivious transfer and secure identification in
  the noisy-quantum-storage model, 2010.
\newblock 1002.1495.

\bibitem{CM97}
Christian Cachin and Ueli~M. Maurer.
\newblock Unconditional security against memory-bounded adversaries.
\newblock In {\em CRYPTO '97: Proceedings of the 17th Annual International
  Cryptology Conference on Advances in Cryptology}, 1997.

\bibitem{qphGC01}
Daniel Gottesman and Isaac~L. Chuang.
\newblock Quantum digital signatures, 2001.
\newblock quant-ph/0105032.

\bibitem{MvOV96}
A.~J~. Menezes, P.~van Oorschot, and S.~Vanstone.
\newblock {\em Handbook of Applied Cryptography}.
\newblock CRC Press LLC, Boca Raton, 1996.

\bibitem{KKNY05}
Akinori Kawachi, Takeshi Koshiba, Harumichi Nishimura, and Tomoyuki
Yamakami.
\newblock Computational indistinguishability between quantum states and its
  cryptographic application.
\newblock In {\em Advances in Cryptology – EUROCRYPT 2005}, volume 3494 of {\em
  Lecture Notes in Computer Science}, pages 268--284. Springer, 2005.
\newblock full version at http://arxiv.org/abs/quant-ph/0403069.

\bibitem{HKK08}
Masahito Hayashi, Akinori Kawachi, and Hirotada Kobayashi.
\newblock Quantum measurements for hidden subgroup problems with optimal sample
  complexity.
\newblock {\em Quantum Information and Computation}, 8:0345--0358, 2008.

\bibitem{Got05}
Daniel Gottesman.
\newblock Quantum public key cryptography with information-theoretic security.
\newblock Workshop on classical and quantum information security, Caltech, 15 -
  18 December, 2005.
\newblock http://www.cpi.caltech.edu/quantum-security/program.html; see also
  http://www.perimeterinstitute.ca/personal/dgottesman.

\bibitem{Kak06}
Subhash Kak.
\newblock A three-stage quantum cryptography protocol.
\newblock {\em Foundations of Physics Letters}, 19:293--296, 2006.

\bibitem{Nik08}
Georgios~M. Nikolopoulos.
\newblock Applications of single-qubit rotations in quantum public-key
  cryptography.
\newblock {\em Phys. Rev. A}, 77:032348, 2008.
\newblock see also Phys. Rev. A. 78, 019903.

\bibitem{BRST06}
Stephen~D. Bartlett, Terry Rudolph, Robert~W. Spekkens, and Peter~S.
Turner.
\newblock Degradation of a quantum reference frame.
\newblock {\em New J. Phys.}, 8:58, 2006.

\bibitem{OTU00}
T.~Okamoto, K.~Tanaka, and S.~Uchiyama.
\newblock Quantum public-key cryptosystems.
\newblock In {\em Proc. of CRYPTO 2000}, volume 1880 of {\em Lecture Notes in
  Computer Science}, pages 147--165. Springer-Verlag, 2000.

\bibitem{Gol01}
O.~Goldreich.
\newblock {\em Foundations of cryptography ({V}olume I): {B}asic tools}.
\newblock Cambridge University Press, Cambridge, 2001.

\bibitem{JUW09}
Rahul Jain, Sarvagya Upadhyay, and John Watrous.
\newblock Two-message quantum interactive proofs are in {PSPACE}.
\newblock {\em Foundations of Computer Science, Annual IEEE Symposium on},
  0:534--543, 2009.

\bibitem{MS07}
Rajat Mittal and Mario Szegedy.
\newblock Product rules in semidefinite programming.
\newblock In Erzsébet Csuhaj-Varjú and Zoltán Ésik, editors, {\em FCT}, volume
  4639 of {\em Lecture Notes in Computer Science}, pages 435--445. Springer,
  2007.

\bibitem{vDdAEMM07}
Wim van Dam, G.~Mauro D'Ariano, Artur Ekert, Chiara Macchiavello,
and Michele
  Mosca.
\newblock Optimal phase estimation in quantum networks.
\newblock {\em Journal of Physics A: Mathematical and Theoretical},
  40:7971--7984, 2007.

\bibitem{BRS03}
Stephen~D. Bartlett, Terry Rudolph, and R.~W. Spekkens.
\newblock Classical and quantum communication without a shared reference frame.
\newblock {\em Phys. Rev. Lett.}, 91:027901, 2003.

\bibitem{qphIM08}
Lawrence~M. Ioannou and Michele Mosca.
\newblock Universal quantum computation in a hidden basis.
\newblock to appear in Quantum Information and Computation;
  http://arxiv.org/abs/0810.2780.

\bibitem{Enk05}
S.~J. van Enk.
\newblock Quantifying the resource of sharing a reference frame.
\newblock {\em Phys. Rev. A}, 71:032339, 2005.

\bibitem{BRS07}
Stephen~D. Bartlett, Terry Rudolph, and Robert~W. Spekkens.
\newblock Reference frames, superselection rules, and quantum information.
\newblock {\em Rev. Mod. Phys.}, 79(555), 2007.

\bibitem{Che09}
Donny Cheung.
\newblock {U}npublished notes, 2009.

\bibitem{Wat08}
John Watrous.
\newblock Theory of quantum information.
\newblock Lecture notes for course CS 789, University of Waterloo. Available at
  http://www.cs.uwaterloo.ca/~watrous/, 2008.

\bibitem{Gut09}
Gustav Gutoski.
\newblock {\em Quantum Strategies and Local Operations}.
\newblock PhD thesis, University of Waterloo, 2009.

\bibitem{CSUU06}
Richard Cleve, William Slofstra, Falk Unger, and Sarvagya Upadhyay.
\newblock Strong parallel repetition theorem for quantum {XOR} proof systems,
  2006.
\newblock arXiv:quant-ph/0608146v1.

\end{thebibliography}

\section*{Appendix}

Consider the following non-cryptographic, two-message interactive protocol (or game) between Evelyn and Bobby (neither of whom is considered adversarial, hence we distinguish these two players from Eve and Bob), denoted $\mathcal{L}=\mathcal{L}(\Phi)$, where $\Phi$ is a quantum operation (super-operator) that specifies Evelyn's action in Step 2' below (the quantities $r$ and $t$ are as defined previously):

\begin{itemize}

\item (1') Bobby chooses a uniformly random $x\in \{1,2,\ldots,2r+1\}$ and
creates a $(t+3)$-qubit system in the state $\myket{\psi_x}^{\otimes
(t+1)}(\sys{\myket{0}\myket{1} + \myket{1}\myket{0}})$; Bobby sends
to Evelyn $t$ copies of $\myket{\psi_x}$ as well as one qubit of the
system in the state $\myket{0}\myket{1} + \myket{1}\myket{0}$.

\item (2') Evelyn carries out the quantum operation $\Phi$ on the received qubits which outputs one qubit, which Evelyn sends to Bobby.

\item (3') Bobby measures the received qubit in the computational basis $\{\myket{0}, \myket{1}\}$; if the measurement outcome is ``1'', then he applies the Pauli-$Z$
gate
\begin{eqnarray}
Z := \left[\begin{array}{cc} 1 & 0 \\ 0 & -1
\end{array}\right]
\end{eqnarray}
to the qubit of the system in the state $\myket{0}\myket{1} + \myket{1}\myket{0}$ that he kept in Step 1'.  Finally, Bobby
\textsc{swap}-tests this qubit with his remaining
copy of $\sys{\myket{\psi_x}}$.
\end{itemize}

The following proposition is immediate:

\begin{prop} The probability that Eve causes Bob's \textsc{swap}-test to pass in any particular iteration of the protocol in Section \ref{sec_ProtSpec} is at most
\begin{eqnarray}
\alpha := \max_{\Phi} \textrm{\emph{Pr}}[\textrm{\emph{Bobby's \textsc{swap}-test passes in }}\mathcal{L}(\Phi)],
\end{eqnarray}
where $\Phi$ ranges over all admissible quantum operations that Evelyn can apply in Step 2'.
\end{prop}

Now consider the parallel $s$-fold repetition of $\mathcal{L}$, which we denote $\mathcal{L}^{\| s} = \mathcal{L}^{\| s} (\Phi')$, where now $\Phi'$ is Evelyn's quantum operation in the second step of $\mathcal{L}^{\| s}$.  The following proposition is also immediate:

\begin{prop}  The probability that Eve fools Bob on the first attempt using $t$ copies in the protocol in Section \ref{sec_ProtSpec} is at most
\begin{eqnarray}
\alpha':= \max_{\Phi'} \textrm{\emph{Pr}}[\textrm{\emph{all of Bobby's \textsc{swap}-tests pass in }}\mathcal{L}^{\| s}(\Phi')],
\end{eqnarray}
where $\Phi'$ ranges over all admissible quantum operations that Evelyn can apply in the second step of $\mathcal{L}^{\| s}$.
\end{prop}

 Therefore, in order to prove that it is sufficient to consider individual (as opposed to coherent) attacks by Eve, it suffices to show that $\alpha' = \alpha^s$.

 In Ref. \cite{JUW09}, it is shown that the maximum acceptance probability of any two-message interactive proof system can be expressed as a semidefinite (optimization) program (see Ref. \cite{Wat08} for a relevant review of semidefinite programming).  Before we apply this fact, we need to make some definitions. Let $\X$ and $\Y$ be the input and output spaces, respectively, of Evelyn's quantum operation $\Phi$ in $\mathcal{L}$, i.e. $\Phi: \lin{\X} \rightarrow \lin{\Y}$, where $\lin{\X}$ is the space of all linear operators from the complex Euclidean space $\X$ to itself (and likewise for $\lin{\Y}$). Let $\pos{\Y \otimes \X}$ denote the set of all positive semidefinite operators in $\lin{\Y \otimes \X}$.  Similarly, for $\mathcal{L}^{\| s}$, we have that $\Phi': \lin{\X^{\otimes s}} \rightarrow \lin{\Y^{\otimes s}}$.  Viewing Bobby's \textsc{swap}-test passing as ``acceptance'' in an interactive proof system, we thus have, according to Ref. \cite{JUW09}, that $\alpha$ and $\alpha'$ can be expressed, respectively, as solutions to the following semidefinite programs $\pi_{\alpha}$ and $\pi_{\alpha'}$:
\begin{center}
  \begin{minipage}{3in}
    \centerline{\underline{$\pi_\alpha$}}\vspace{-7mm}
    \begin{eqnarray*}
    \text{maximize:} && \tr(B^\dagger X)\\
    \text{subject to:} &&
    \tr_\Y (X) = \I_{\X},\\
    && X\in\pos{ \Y \otimes \X}
    \end{eqnarray*}
  \end{minipage}
  \begin{minipage}{3in}
    \centerline{\underline{$\pi_{\alpha'}$}}\vspace{-7mm}
    \begin{eqnarray*}
    \text{maximize:} && \tr((B^{\otimes s})^\dagger X')\\
    \text{subject to:} &&
    \tr_{\Y^{\otimes s}}(X') = I_{\X^{\otimes s}},\\
    && X'\in\pos{ (\Y \otimes \X)^{\otimes s}},
    \end{eqnarray*}
  \end{minipage}
\end{center}
where $X$ and $X'$ are the Choi-Jamio{\l}kowski representations of $\Phi$ and $\Phi'$, and $B$ is a positive semidefinite operator representing Bobby's actions, i.e., $B \in \pos{\Y \otimes \X}$.  Furthermore, it is shown in Ref. \cite{JUW09} that such semidefinite programs (arising from two-message interactive proof systems) satisfy the condition of strong duality, which means that the solution to each semidefinite program above coincides with that of its dual.

In Ref. \cite{MS07}, the following theorem is proven:

\begin{theorem}[\cite{MS07}]\label{thm_MittalSzegedy}
Suppose that the following two semidefinite programs $\pi_1$ and $\pi_2$ satisfy strong duality:
\begin{center}
  \begin{minipage}{3in}
    \centerline{\underline{$\pi_1$}}\vspace{-7mm}
    \begin{eqnarray*}
    \text{maximize:} && \textrm{\emph{Tr}}(J_1^\dagger X)\\
    \text{subject to:} &&
    \Phi_1 (X) = C_1 ,\\
    && X\in\pos{\X_1}
    \end{eqnarray*}
  \end{minipage}
  \begin{minipage}{3in}
    \centerline{\underline{$\pi_2$}}\vspace{-7mm}
    \begin{eqnarray*}
    \text{maximize:} && \textrm{\emph{Tr}}(J_2^\dagger X)\\
    \text{subject to:} &&
    \Phi_2 (X) = C_2 ,\\
    && X\in\pos{\X_2},
    \end{eqnarray*}
  \end{minipage}
\end{center}
where $\Phi_1: \lin{\X_1} \rightarrow \lin{\Y_1}$ and $\Phi_2: \lin{\X_2} \rightarrow \lin{\Y_2}$, for complex Euclidean spaces $\X_1, \Y_1, \X_2, \Y_2$, and $J_1 \in \lin{\X_1}$ and $J_2\in\lin{\X_2}$ are Hermitian.  Let $\alpha(\pi_1)$ and $\alpha(\pi_2)$ denote the semidefinite programs' solutions.  If $J_1$ and $J_2$ are positive semidefinite, then the solution to the following semidefinite program, denoted $\pi_1 \otimes \pi_2$, is $\alpha(\pi_1 \otimes \pi_2) = \alpha(\pi_1) \alpha(\pi_2)$:
\begin{center}
  \begin{minipage}{3in}
    \centerline{\underline{$\pi_1 \otimes \pi_2$}}\vspace{-7mm}
    \begin{eqnarray*}
    \text{maximize:} && \textrm{\emph{Tr}}( (J_1\otimes J_2)^\dagger X)\\
    \text{subject to:} &&
    \Phi_{1} \otimes \Phi_2 (X) = C_1 \otimes C_2,\\
    && X\in\pos{\X_1 \otimes \X_2}.
    \end{eqnarray*}
  \end{minipage}
\end{center}
\end{theorem}

Since $B$ is positive semidefinite and $\pi_{\alpha'} = \pi_\alpha^{\otimes s}$ (using the associativity of $\otimes$), Theorem \ref{thm_MittalSzegedy} can be applied $s-1$ times in order to prove that $\alpha' = \alpha^s$ as required.  See Ref. \cite{Gut09} for a similar approach, based on ideas in Ref. \cite{CSUU06}.

  Note that this argument, combined with the arguments in the main body of the paper, shows that both the serial and parallel versions of our identification protocol are secure.
\end{document}